\documentclass{elsart}
\usepackage{english}
\newcommand\Tr{{\rm Tr}}
\newcommand\nn{\nonumber\\}
\newcommand{\fr}[2]{{\frac{#1}{#2}}}
\newcommand{\la}[1]{\label{#1}}
\newcommand{\be}{\begin{equation}}
\newcommand{\ee}{\end{equation}}
\newcommand{\ba}{\begin{eqnarray}}
\newcommand{\ea}{\end{eqnarray}}
\newcommand{\nr}[1]{(\ref{#1})}
\newcommand{\msbar}{\overline{\mbox{\rm MS}}}
\def\undertilde#1{\mathop{\vtop{\ialign{##\cr$\textstyle{#1}$\cr%
\noalign{\kern1pt\nointerlineskip}\hfil$\mathchar"0365$\hfil\cr}}}}
\def\wideundertilde#1{\mathop{\vtop{\ialign{##\cr$\textstyle{#1}$\cr%
\noalign{\kern1pt\nointerlineskip}\hfil$\mathchar"0367$\hfil\cr}}}}
\begin{document}
\begin{frontmatter}
\begin{flushright}
HD-THEP-97-16, hep-lat/9705003
\end{flushright}
\title{Lattice-continuum relations for \\
3d SU(N)+Higgs theories}
\author{M. Laine\thanksref{mikko}}
\address{Institut f\"ur Theoretische Physik, 
Philosophenweg 16,\\ 
D-69120 Heidelberg, Germany}
\author{A. Rajantie\thanksref{arttu}}
\address{Department of Physics, P.O. Box 9,\\
FIN-00014 University of Helsinki, Finland}
\thanks[mikko]{m.laine@thphys.uni-heidelberg.de}
\thanks[arttu]{arttu.rajantie@helsinki.fi}
\maketitle
\abstract
3d lattice studies have recently attracted a lot of attention, 
especially in connection with finite temperature field theories.
One ingredient in these studies is a perturbative
computation of the 2-loop lattice counterterms, which are exact
in the continuum limit. We extend previous such results to SU($N$) 
gauge theories with Higgs fields in the fundamental and adjoint 
representations. The fundamental SU(3)$\times$SU(2) case
might be relevant for the electroweak phase transition in the MSSM,
and the adjoint case for the GUT phase transition and for
QCD in the high temperature phase. 
We also revisit the standard SU(2)$\times$U(1) and U(1) theories.

\endabstract
\end{frontmatter}

\section{Introduction}

Finite temperature gauge theories exhibit many interesting and
important phenomena, such as the EW (electroweak) and QCD phase
transitions. Unfortunately these are not easy questions to answer,
due in part to the infrared problem at finite temperature. However, 
recently a successful approach has been developed at least for the
thermodynamical aspects of the EW phase transition 
(for a review, see~\cite{rs}). This approach
consists of a perturbative dimensional reduction into a 3d effective
theory~\cite{g,generic,jkp,br,mssm,ref:su5} 
and of non-perturbative lattice simulations in that 
theory~\cite{krs,su2u1,leip,karsch,phtw,mt}.
Other cases where similar 3d  simulations are needed are
QCD in the high temperature phase~\cite{qcd} and
the Ginzburg-Landau model of superconductivity~\cite{u1,farakos}.

One ingredient in the non-perturbative 3d studies is the 
relation of the lattice and continuum regularization 
schemes~\cite{ref:hpert1,ref:hpert2}.
This problem is analogous to the determination of 
$\Lambda_{\overline{\rm MS}}/\Lambda_{\rm latt}$ in 4d~\cite{hh}.
In this paper, we extend previous 3d results to theories
relevant for extensions of the Standard Model. 

To be more specific, consider that 
one has to convert the results of lattice simulations to
continuum observables. In 4d this is 
usually done by measuring some
experimentally known observables and using them to fix the 
parameters.
However, in 3d this is not possible 
since no such observables exist. 
Thus one has to use
lattice perturbation theory \cite{ref:rothe}
to explicitly renormalize the theory and find the relationship to
the $\overline{\rm MS}$ scheme observables in this way. For 
super-renormalizable theories the result will be exact in the
continuum limit at a finite order in the loop expansion; 
in 3d one needs to go to 2 loops, and only 
the mass parameter gets renormalized.

The task of relating lattice and continuum
observables for 3d SU(2)+Higgs 
theories was previously
discussed in \cite{ref:hpert1,ref:hpert2}. In \cite{ref:hpert2}
the calculations were shown in some detail and results were given
for SU(2) and U(1) theories with Higgs fields in adjoint and fundamental
representations. Here we go to SU($N$) gauge fields with 
both types of scalar fields.
1-loop ${\mathcal O}(a)$-improvement in 3d has been 
discussed in~\cite{moore}.

It should be noted that 
for some observables the 2-loop mass counterterms are not needed. 
Such case is, for instance, the
determination of the discontinuity in
$\langle\phi^\dagger\phi\rangle/g_3^2$
at the phase transition point ($m_c^2$)
in a theory with only one mass parameter, 
if one does not need to know the 
numerical value of $m_c^2$~\cite{moore}.
However, in most cases the value of $m_c^2$ is needed:
one needs it in order to make a systematic comparison 
with 3d perturbation theory 
(the lattice and perturbative results should be compared 
for the same parameters), one needs it if there are
several mass parameters in the theory 
as in the SU(3)$\times$SU(2) model
discussed in Sec.~\ref{sec:other}, and one needs 
it when connection to 4d physics is made,
especially in theories
sensitive to the value of 
$T/\Lambda_{\overline{\rm MS}}$ such as
QCD in the high-temperature phase~\cite{qcd}. 

This paper is organized as follows. 
In Sec.~\ref{sec:calc} we briefly discuss some general aspects
of the problem. In Sec.~\ref{sec:adjoint} we calculate the
2-loop lattice counterterms for the SU($N$)+adjoint Higgs 
theory and in Sec.~\ref{sec:fund} for the SU($N$)+fundamental Higgs 
theory. Some other fundamental theories, namely SU(3)$\times$SU(2)
with two Higgses and SU(2)$\times$U(1), U(1) with one 
Higgs, are discussed in Sec.~\ref{sec:other}. 
In Appendix A we discuss the Feynman rules, 
in App.~B some lattice integrals, 
in App.~C the isospin contractions, 
in App.~D the numerical values of the constants appearing, 
and in App.~E a unified representation for the $g^2,g^4$-parts
of the counterterms.

\section{Calculations}
\label{sec:calc}

In order to find the relationship between
the lattice and $\overline{\rm MS}$
renormalization schemes we have to calculate some physical quantity in
both schemes and equate them. As discussed in \cite{ref:hpert2}, the
simplest choice is the minimum value of the effective potential.
Practically this means requiring that the $\phi^2$-parts of
the effective potentials coincide. 

In addition to the $\phi^2$-term, one needs to  compute
the mass-dependent part of the vacuum counterterm.
Its definition is that if it is added to the lattice
Lagrangian, 
${\mathcal L}_{\rm lat}\to {\mathcal L}_{\rm lat}+\delta V$, 
then the mass-dependent part of the vacuum energy density agrees with 
the $\msbar$-result. The significance of the
vacuum counterterm is that
if ${\mathcal L}_{\rm latt}=m^2\phi^\dagger\phi+\ldots$, then  
\be
\langle\phi^\dagger\phi\rangle=\frac{d V({\rm min})}{d m^2}.
\ee
Hence, the relation of $\langle\phi^\dagger\phi\rangle$ measured in 
continuum and on the lattice {\it without} adding
$\delta V$ to the Lagrangian is 
\be
\langle\phi^\dagger\phi\rangle_{\rm cont} = 
\langle\phi^\dagger\phi\rangle_{\rm latt} + 
\frac{d(\delta V)}{d m^2}.
\ee
This equation can be used to infer the non-perturbative value of 
$\langle\phi^\dagger\phi\rangle_{\rm cont}$ from 
lattice simulations. 
Like for the mass counterterm, the 2-loop result 
is exact in the continuum limit.
If one has a phase transition 
and is only interested in the discontinuity of
$\langle\phi^\dagger\phi\rangle$, then the vacuum 
counterterm does not contribute. 

When replacing SU(2) with SU($N$), the Feynman rules
discussed in~\cite{ref:hpert2} get modified and
obtain a more complicated form. In most cases this means only
replacing $\epsilon^{ABC}$ with $f^{ABC}$, but sometimes more
structure is added. The most crucial difference is that 
the symmetric tensors $d^{ABC}$ are no longer zero.
The Feynman rules for the gauge and ghost-gauge vertices
can be read from \cite{ref:rothe},
with the replacements $1/4a^2\rightarrow N/12a$ in Eq. (14.39),
and $(2/3)(\delta^{AB}\delta^{CD}+\ldots)\rightarrow 
(2/N)(\delta^{AB}\delta^{CD}+\ldots)$ in Eq. (14.44). 
The sign of the 
$\overline{c}cAA$-vertex in \cite{ref:rothe} should 
also be reversed:
the ghost--gluon part of the Lagrangian is
\ba
&& S_{\rm ghost} = 
\overline{c}^A(p)c^A(p)\widetilde{p}^2 +
ig f^{ABC} \overline{c}^A(p)A^B_i(q)c^C(r)\undertilde{r_i}\widetilde{p}_i
\nn
&& \quad -\frac{1}{24} g^2a^2
\Bigl(f^{ACE}f^{BDE}+f^{ADE}f^{BCE}\Bigr) 
\overline{c}^A(p)A^C_i(q)A^D_i(r)c^B(s) \widetilde{s}_i
\widetilde{p}_i,
\ea
where momentum conservation and 
due integrations are implied, and 
$\widetilde{p}_i=\frac{2}{a}\sin\frac{a}{2}p_i$, 
$\undertilde{r_i}=\cos\frac{a}{2}r_i$.
The other vertices are given in Appendix~\ref{app:verteksit}.

We will perform the calculations in the $R_\xi$ gauge with $\xi=1$
in order to make the gauge propagator simple --
the result is gauge fixing independent. 
As mentioned, we only need to go to 2-loop level to 
obtain the exact result. 
In \cite{ref:hpert2} the results for
most of the necessary integrals were given.
In the present case, different components of the
fields get different masses in the symmetry breakdown and
we have to consider the corresponding integrals.
However, it is easy to express all these new integrals using the
ones defined in \cite{ref:hpert2}; the most important cases are
for completeness
given in Appendix~\ref{app:ints}.

\section{Adjoint Higgs}
\la{sec:adjoint}

Let us consider first a system with an SU($N$) gauge field and a Higgs
scalar field in the adjoint representation of the gauge group.
The Lagrangian on the lattice is
\begin{eqnarray}
{\mathcal L}_{\rm latt}
&=&
\frac{1}{a^4g^2}\sum_{i,j}\Tr\left[{\bf 1}-P_{ij}(x)\right]\nn
&&+
\frac{2}{a^2}\sum_i
\left[\Tr\Phi(x)^2-\Tr\Phi(x)U_i(x)\Phi(x+i)U_i^{-1}(x)\right]\nn
&&+
m^2\Tr\Phi^2+\lambda_1(\Tr\Phi^2)^2+
\lambda_2\Tr\Phi^4,
\end{eqnarray}
where $U_i(x)=\exp[iagA_i(x)]$,
$A_i=A^A_iT^A$, $x+i\equiv  x+a\pol\e_i$,
\begin{equation}
P_{ij}(x)=U_i(x)U_j(x+i)U_i^{-1}(x+j)U_j^{-1}(x),
\end{equation}
and $a$ is the lattice constant.

When $a\rightarrow 0$, this becomes the usual continuum Lagrangian
\begin{equation}
{\mathcal L}_{\rm cont}
=\frac{1}{2}\Tr F_{ij}F_{ij}+\Tr D_i\Phi D_i\Phi+m^2\Tr\Phi^2
+\lambda_1(\Tr\Phi^2)^2+
\lambda_2\Tr\Phi^4,
\end{equation}
where $F_{ij}=\partial_iA_j-\partial_jA_i+ig[A_i,A_j]$ and 
$D_i\Phi=\partial_i\Phi+ig[A_i,\Phi]$.

We can write
$\Phi=\sum_{A=1}^{N^2-1}\Phi^AT^A,$
where the generators $T^A$ 
of the Lie algebra of the group SU($N$)
are chosen to be Hermitian 
$N\times N$ matrices and
are normalized as $\Tr T^AT^B=\frac{1}{2}\delta^{AB}.$ 
If $\hat\phi$ is a real $N$-dimensional vector, then
\begin{equation}
\label{eq:tn2-1}
T^{N^2-1}=\frac{1}{\sqrt{2N(N-1)}}\left({\bf 1}-N
\frac{\hat\phi\hat{\phi}^\dagger}{\hat{\phi}^\dagger\hat\phi}\right)
\end{equation}
is Hermitian, traceless and properly normalized and can therefore
be chosen to be one of the
generators. 

We wish to relate the lattice observables to the $\overline{\rm MS}$
scheme ones using the method given in
\cite{ref:hpert1,ref:hpert2}. 
In order to calculate the effective potential, we break the 
symmetry by shifting 
the Higgs field $\Phi\rightarrow\Phi+vT^{N^2-1}$. 
The effective potential in the adjoint case depends on the
direction of the shift. However, we are only interested in the
quadratic part of the potential which depends only on the magnitude $v$
and thus any direction can be chosen.
The Higgs-gauge cross term is cancelled by introducing an $R_\xi$ gauge 
fixing term
${\mathcal L}_\xi=F^AF^A/2\xi a^2$, where
\begin{equation}
F^A(x)=\sum_i\left[A_i^A(x)-A_i^A(x-i)\right]
+\xi agvf^{A,B,N^2-1}\Phi^B(x).
\end{equation}

If we define the projection operators
\ba
\label{eq:projadj}
&& P_1^{AB} = 
\delta^{AB} - P_2^{AB} - P_3^{AB};
\quad P_1^{AA}=N(N-2), \nn
&& P_2^{AB} = 
-4\frac{\hat{\phi}^\dagger T^A\hat{\phi}}{\hat{\phi}^\dagger\hat{\phi}}
\frac{\hat{\phi}^\dagger T^B\hat{\phi}}{\hat{\phi}^\dagger\hat{\phi}}
+2  \frac{\hat{\phi}^\dagger T^{\{A}T^{B\}}
\hat{\phi}}{\hat{\phi}^\dagger\hat{\phi}}; 
\quad P_2^{AA}=2(N-1), \nn
&& P_3^{AB} =  
\frac{2N}{N-1}\frac{\hat{\phi}^\dagger 
T^A\hat{\phi}}{\hat{\phi}^\dagger\hat{\phi}}
\frac{\hat{\phi}^\dagger T^B\hat{\phi}}{\hat{\phi}^\dagger\hat{\phi}};
\quad P_3^{AA}=1,
\ea
we can write the propagators in the broken phase as
\ba
&&\langle \Phi^A \Phi^B\rangle =
\biggl(\frac{P_1^{AB}}{\widetilde{k}^2+m_1^2}+
\frac{P_2^{AB}}{\widetilde{k}^2+m_2^2}+
\frac{P_3^{AB}}{\widetilde{k}^2+m_3^2}
\biggr),\nn
&&\langle A_i^A A_j^B\rangle =\delta_{ij}
\biggl(\frac{P_1^{AB}}{\widetilde{k}^2}+
\frac{P_2^{AB}}{\widetilde{k}^2+M^2}+
\frac{P_3^{AB}}{\widetilde{k}^2}
\biggr), \quad
\langle \overline{c}^A c^B \rangle = - \biggl(\ldots\biggr),
\ea
where $(\ldots)$ is the same expression as in the gauge propagator and
the masses are
\begin{eqnarray}
\label{eq:masses}
M^2&=&\frac{N}{2(N-1)}g^2v^2,\nn
m_1^2&=&m^2+
\left(\lambda_1+\frac{3}{N(N-1)}\lambda_2\right)v^2,\nn
m_2^2&=&m^2+
\left(\lambda_1+\frac{N^2-3N+3}{N(N-1)}\lambda_2\right)v^2+M^2,\nn
m_3^2&=&m^2+
3\left(\lambda_1+\frac{N^2-3N+3}{N(N-1)}\lambda_2\right)v^2.
\end{eqnarray}

It is now straightforward to calculate the renormalization counterterms
and to relate them to the ones obtained in the $\overline{\rm MS}$ scheme.
Let $m^2(\mu)$ be the renormalized mass in the $\overline{\rm MS}$ scheme
and $m^2=m^2(\mu)+\delta m^2(\mu)$ the bare mass. 
The diagrams needed to calculate $\delta m^2$ are the same as in
\cite{ref:hpert2}. 
The Feynman rules for the vertices are given in Appendix \ref{app:verteksit}.
The calculation of the isospin factors in the broken
phase
is discussed in Appendix \ref{app:const}, and some typical
lattice integrals are in Appendix~\ref{app:ints}. 
For some diagrams, products
of four structure constants are needed. Their values are given in
Appendix A of~\cite{ref:su5}.

Adding the contributions from the different diagrams together we get
\begin{eqnarray}
\label{eq:deltam}
\delta m^2&=&
-\left(2Ng^2+(N^2+1)\lambda_1+(2N^2-3)\frac{\lambda_2}{N}\right)
\frac{\Sigma}{4\pi a}\nonumber\\
&&-\frac{1}{16\pi^2}\left\{
\left[
2Ng^2\left((N^2+1)\lambda_1+(2N^2-3)\frac{\lambda_2}{N}\right)
-2(N^2+1)\lambda_1^2\right.\right.\nonumber\\
&&\left.\left.-4(2N^2-3)\lambda_1\frac{\lambda_2}{N}
-(N^4-6N^2+18)\frac{\lambda_2^2}{N^2}\right]
\left(\ln\frac{6}{a\mu}+\zeta\right)\right.\nonumber\\
&&\left.
+2Ng^2\left((N^2+1)\lambda_1+(2N^2-3)\frac{\lambda_2}{N}\right)
\left(\frac{1}{4}\Sigma^2-\delta\right)\right.\nonumber\\
&&\left.
+g^4N^2\left[
\frac{5}{8}\Sigma^2+\left(\frac{1}{2}-\frac{4}{3N^2}\right)\pi\Sigma
-4(\delta+\rho)+2\kappa_1-\kappa_4
\right]
\right\},
\end{eqnarray}
where the constants
$\zeta$, $\delta$, $\rho$, $\kappa_1$, $\kappa_4$ and $\Sigma$
have been defined in \cite{ref:hpert1,ref:hpert2}. Their numerical
values are given in Appendix \ref{app:num}.
The $\mu$ dependences of $\delta m^2$ and $m^2$ cancel as they should,
which can be seen by
comparison with Eq. (48) of \cite{ref:su5}.
As for SU(2), the finite parts of $\delta_{ii}/\epsilon$ arising from
dimensional regularization in the $\overline{\rm MS}$ case, cancel
between the diagrams (vvv) and (vvs). The 2-loop $1/a$-terms
cancel against contributions from the mass counterterms.

The mass-dependent part of the vacuum counterterm
is also straightforward to calculate, and it implies that
\ba
\Bigl\langle\Tr \Phi^2 \Bigr\rangle_{\rm cont} & = & 
\Bigl\langle\Tr \Phi^2 \Bigr\rangle_{\rm latt}
-({N^2-1})\frac{\Sigma}{8\pi a} \nn
&& -\frac{g^2}{16\pi^2}N(N^2-1)
\left(\ln\frac{6}{a\mu}+\zeta+\frac{\Sigma^2}{4}-\delta\right).
\la{adjpdp}
\ea

As a concrete example, consider SU(5).
Using the numerical values of the constants 
$\Sigma,\delta,\rho,\zeta,\kappa_1,\kappa_4$
given in Appendix~\ref{app:num}, one gets
from  eq.~\nr{eq:deltam}, 
\ba
m^2&=& m^2(\mu)
-\left(10 g^2+26 \lambda_1+\frac{47}{5}\lambda_2\right)
\frac{\Sigma}{4\pi a}\nonumber\\
&&-\frac{1}{16\pi^2}\biggl\{
\left[
 10 g^2\left( 26 \lambda_1+\frac{47}{5}\lambda_2\right)
-52\lambda_1^2 
-\frac{188}{5}\lambda_1\lambda_2
-\frac{493}{25}\lambda_2^2\right]
\nonumber\\
&&\times \biggl(\ln\frac{6}{a\mu}+\zeta\biggr)
+124.01g^4 +5.7947g^2\left(26 \lambda_1+\frac{47}{5}\lambda_2\right)
\biggr\}.
\la{eq:su5num}
\ea
The combination appearing on the last row in eq.~\nr{adjpdp}
is in eq.~\nr{eq:pdpnum}.

\section{Fundamental Higgs}
\la{sec:fund}

For the fundamental theory, the lattice Lagrangian is
\ba
{\mathcal L}_{\rm lat} & = & \frac{1}{a^4 g^2}\sum_{i,j} {\rm Tr}
\Bigl[{\bf 1}-P_{ij}(x)\Bigr]\nonumber \\
& +  & \frac{2}{a^2}\sum_i\Bigl[
\phi^\dagger(x)\phi(x)-
{\rm Re}\,\phi^\dagger(x)U_i(x)\phi(x+i)\Bigr]\nonumber \\ 
& + &
m^2\phi^\dagger\phi+
\lambda (\phi^\dagger\phi)^2.
\label{eq:fundlL}
\ea
The corresponding continuum theory is 
\be
{\mathcal L}_{\rm cont} = 
\fr12 \Tr  F_{ij}F_{ij}
+(D_i \phi)^\dagger(D_i \phi)+m^2 \phi^\dagger \phi+
\lambda (\phi^\dagger \phi)^2,
\ee
where $F_{ij}=T^A F_{ij}^A$ and
$D_i=\partial_i+i g T^A A_i^A$.

In contrary to the adjoint case, 
the effective potential 
can be calculated 
in the fundamental case 
for an arbitrary shift (see, e.g., \cite{bjls}).
We make a shift $\hat\phi$ such that
\be
\hat{\phi}^\dagger\hat{\phi}=\frac{v^2}{2}.
\ee
For simplicity, we shall assume $\hat\phi$ to be real
(this assumption enters in the scalar propagators). 
The gauge fixing term is 
\be
F^A = \sum_i [A_i^A(x)-A_i^A(x-i)] + i \xi a g  
\Bigl(\phi^\dagger T^A\hat{\phi}-\hat{\phi}^\dagger T^A \phi\Bigr).
\ee
As mentioned, in practical calculations
we use $\xi=1$ to simplify the momentum algebra.
The shifted field $\tilde\phi=\phi-\hat\phi$
is written as
$\tilde{\phi}_\alpha = 
(u_\alpha+i\omega_\alpha)/\sqrt{2}$, $\alpha=1,\ldots ,N$. 
To get the
propagators we define two more projection operators
in addition to those in eq.~\nr{eq:projadj},
\ba
&&  T_{\alpha\beta}=\delta_{\alpha\beta}-
\frac{\hat{\phi}_\alpha\hat{\phi}_\beta}{\hat{\phi}^\dagger\hat{\phi}},\quad
L_{\alpha\beta}=
\frac{\hat{\phi}_\alpha\hat{\phi}_\beta}{\hat{\phi}^\dagger\hat{\phi}};
\qquad  T_{\alpha\alpha}=N-1,\quad L_{\alpha\alpha}=1.
\ea
Then the propagators are
\ba
&& \langle u_\alpha u_\beta\rangle =  
\frac{T_{\alpha\beta}}{\widetilde{k}^2+
m_\omega^2}+ \frac{L_{\alpha\beta}}{\widetilde{k}^2+m_u^2}, \quad
\langle\omega_\alpha\omega_\beta\rangle =  
\frac{T_{\alpha\beta}}{\widetilde{k}^2+
m_\omega^2}+ 
\frac{L_{\alpha\beta}}{\widetilde{k}^2+\overline{m}_\omega^2}, \nn
&& \langle A^A_i A^B_j\rangle =
\delta_{ij}
\biggl(\frac{P_1^{AB}}{\widetilde{k}^2}+
\frac{P_2^{AB}}{\widetilde{k}^2+M^2}+
\frac{P_3^{AB}}{\widetilde{k}^2+\overline{M}^2}
\biggr),\quad  
\langle \overline{c}^A c^B \rangle = - \biggl(\ldots\biggr).
\ea
Here $\overline{c}^A, c^B$ are the ghost fields,
the expression $(\ldots)$ is the same as for the gauge fields, 
and the non-vanishing masses are
\ba
M^2 & = & \fr14 g^2 v^2,\quad
\overline{M}^2=\frac{2(N-1)}{N}M^2, \nn
m_\omega^2 & =  & m^2+\lambda v^2+ M^2, \quad
\overline{m}_\omega^2 = 
m^2+\lambda v^2+ \overline{M}^2, \nn
m_u^2 & = &  m^2+3\lambda v^2.
\la{eq:fmasses}
\ea

The isospin algebra appearing in the graphs is 
for the most part the same as in continuum, 
and can be read from~\cite{bjls}.
There are a few additional terms which do not appear in 
continuum since the vertex is proportional to powers of
$a^2$. Such terms come from the $\overline{c} c A A$-vertex 
and the two $A A A A$-vertices. The isospin algebra 
of $\langle \overline{c} c A A\rangle$ can be deduced 
from the graph (vv) in eq.~(B.46) in~\cite{bjls}. 
The algebra related to 
the $AAAA$-vertex in eq.~\nr{fundAAAA} is
\ba
&&
\Bigl(\delta^{AB}\delta^{CD}+\delta^{AC}\delta^{BD}+
\delta^{AD}\delta^{BC}\Bigr)
\frac{\phi^\dagger T^AT^BT^CT^D \phi}{\phi^\dagger\phi}=
\frac{(N^2-1)(2N^2-3)}{4N^2}, \nonumber
\ea
where we used that the masses in the propagators can be
put to zero here. The algebra
related to the symmetric $a^4$-part of
the four-gluon vertex coming from the plaquette is  
\ba
&& 
\langle A^A A^B \rangle
\langle A^C A^D\rangle 
\biggl[\biggl(\fr2N 
\delta^{AB}\delta^{CD}+ d^{ABE}d^{CDE}\biggr) + 
(B\leftrightarrow C)+(B\leftrightarrow D) \biggr] \nn
& & \to 
\frac{4}{N}(2N^2-3)\delta^{CD}
\langle A^CA^D\rangle_{M^2-{\rm part}}
=\frac{4(N^2-1)(2N^2-3)}{N^2} g^2\hat{\phi}^\dagger\hat{\phi}.
\la{eq:vvsy}
\ea
Here we used eq.~\nr{eq:vvasyint}
according to which the result of the momentum integral is symmet\-ric and 
quadra\-tic in the masses in the continuum limit. 

After the momentum integrals, 
one can verify the 
cancellation of $1/a$-terms against
1-loop counterterm contributions. 
The result left is 
\ba
m^2 &=& m^2(\mu)
-\left(\frac{N^2-1}{N}g^2+2 (N+1)\lambda\right)
\frac{\Sigma}{4\pi a}\nonumber\\
&&-\frac{1}{16\pi^2}(N^2-1)\biggl\{
\biggl[g^4\frac{4N^2-N+3}{4N^2}+
2\lambda g^2  \frac{N+1}{N}
-4 \lambda^2\frac{1}{N-1}\biggr]
\nonumber\\
&&
\times \biggl(\ln\frac{6}{a\mu}+\zeta\biggr)
+2\lambda g^2 \frac{N+1}{N}
\left(\frac{1}{4}\Sigma^2-\delta\right)
\nonumber\\
&&
+g^4\frac{1}{4 N^2}\biggl[
\frac{4N^2-1}{4}\Sigma^2+\frac{3N^2-8}{3}\pi\Sigma+N^2+1
\nonumber\\
&&
-4N(N+1)\rho-2(3N^2-1)\delta+2N^2(2\kappa_1-\kappa_4)
\biggr]
\biggr\}.
\la{eq:fundres}
\ea
In fact, the 1-loop terms proportional to $g^2$ and the
2-loop terms proportional to $g^4$ in eqs.~\nr{eq:deltam}, \nr{eq:fundres} 
can be written in a unified way, and for completeness this formula is given 
in Appendix~\ref{app:uni}.

The relation of $\langle\phi^\dagger\phi\rangle$ measured in 
continuum and on the lattice is 
\ba
\langle\phi^\dagger\phi\rangle_{\rm cont} & =  & 
\langle\phi^\dagger\phi\rangle_{\rm latt} -\frac{N\Sigma}{4\pi a} \nn
& &  -\frac{1}{(4\pi)^2}g^2(N^2-1)\biggl(
\ln\frac{6}{a\mu}+\zeta+\fr14 \Sigma^2-\delta
\biggr). 
\la{eq:pdp}
\ea
Corrections of order $a$ to the discontinuity of
$\langle\phi^\dagger\phi\rangle$ are
discussed in~\cite{moore}.

\section{Other fundamental Higgs theories}
\la{sec:other}

\subsection{SU(N)$\times$SU(2)}

It has been argued in~\cite{mssm,bjls} that an SU(3)$\times$SU(2)
gauge theory with two scalar fields
might be relevant for the 
electroweak phase transition in MSSM. Let us here consider
for generality SU($N$)$\times$SU(2). The continuum Lagrangian is
\ba
{\mathcal L}_{\rm cont} & = &
\fr14 F^a_{ij}F^a_{ij}+\fr14 G^A_{ij}G^A_{ij} 
+(D_i^w H)^\dagger(D_i^w H)+m_{H}^2 H^\dagger H+
\lambda_{H} (H^\dagger H)^2 \nn 
& + & (D_i^s U)^\dagger(D_i^s U)+m_{U}^2U^\dagger U+
\lambda_{U} (U^\dagger U)^2
+ \gamma H^\dagger H U^\dagger U. \la{Uthe}
\ea
Here $D_i^w=\partial_i+i g_{W}t^a A_i^a$ and
$D_i^s=\partial_i+i g_{S}T^A C_i^A$ are the 
SU(2) and SU($N$) covariant derivatives
($t^a=\sigma^a/2$ where $\sigma^a$ are the Pauli matrices), 
$g_{W}$ and $g_{S}$ are the corresponding gauge couplings, 
$H$ is the Higgs doublet and $U$ is the right-handed stop field. 
Since the two scalar fields are coupled only through 
the simple scalar vertex proportional to $\gamma$, 
discretization follows immediately from Sec.~\ref{sec:fund}.

The results for the counterterms can also mostly be read
from Sec.~\ref{sec:fund}. The reason is that 
when one makes a shift in $H$, say, then in the SU($N$)-sector one only 
needs to replace $m_U^2 \to m_U^2+\gamma \hat H^\dagger \hat H$
in the masses in eq.~\nr{eq:fmasses}. 
Thus the result can be deduced from the mass-dependent part
of the vacuum counterterm.
The only new graph is the scalar figure-8 graph 
proportional to~$\gamma^2$.

The results for the bare lattice mass parameters are
\ba
m_U^2 &=& m_U^2(\mu)
-\left(\frac{N^2-1}{N}g_S^2+2 (N+1)\lambda_U+2\gamma\right)
\frac{\Sigma}{4\pi a}\nonumber\\
&&-\frac{1}{16\pi^2}(N^2-1)\biggl\{
\biggl[g_S^4\frac{4N^2-N+3}{4N^2}+
2\lambda_U g_S^2 \frac{N+1}{N} \nn 
&& +\frac{1}{N^2-1}\biggl(3 \gamma g_W^2
-4 (N+1)\lambda_U^2- 2 \gamma^2\biggr)\biggr]
\left(\ln\frac{6}{a\mu}+\zeta\right)
\nonumber\\
&&
+\biggl(2\lambda_U g_S^2\frac{N+1}{N}+
\frac{3}{N^2-1}\gamma g_W^2 \biggr)
\left(\frac{1}{4}\Sigma^2-\delta\right)
\nonumber\\
&&
+g_S^4\frac{1}{4 N^2}\biggl[
\frac{4N^2-1}{4}\Sigma^2+\frac{3N^2-8}{3}\pi\Sigma+N^2+1
\nonumber\\
&&
-4N(N+1)\rho-2(3N^2-1)\delta+2N^2(2\kappa_1-\kappa_4)
\biggr]
\biggr\}, 
\la{eq:mmU}\\
m_H^2 &=& m_H^2(\mu)
-\left(\frac{3}{2}g_W^2+6\lambda_H+N\gamma\right)
\frac{\Sigma}{4\pi a}\nonumber\\
&&-\frac{1}{16\pi^2}\biggl\{
\left[\frac{51}{16}g_W^4+
9\lambda_H g_W^2 
-12 \lambda_H^2+(N^2-1)\gamma g_S^2-
N\gamma^2\right] \nn
&& \times 
\left(\ln\frac{6}{a\mu}+\zeta\right)
+\biggl[9 \lambda_H g_W^2 +
(N^2-1) \gamma g_S^2 \biggr]
\left(\frac{1}{4}\Sigma^2-\delta\right)
\nonumber\\
&&
+g_W^4\frac{3}{16}\biggl[
\frac{15}{4}\Sigma^2+\frac{4}{3}\pi\Sigma+5
-24\rho-22\delta+8(2\kappa_1-\kappa_4)
\biggr]
\biggr\}.
\la{eq:mmH}
\ea
Using the numerical values in Appendix~\ref{app:num}, 
one gets for $N=3$,
\ba
m_U^2 &=& m_U^2(\mu)
-\left(\frac{8}{3}g_S^2+8 \lambda_U+2\gamma\right)
\frac{\Sigma}{4\pi a}\nonumber\\
&&-\frac{1}{16\pi^2}\biggl\{
\biggl[8 g_S^4+\frac{64}{3} \lambda_U g_S^2 
- 16 \lambda_U^2 + 3 \gamma g_W^2
- 2 \gamma^2\biggr]
\biggl(\ln\frac{6}{a\mu}+\zeta\biggr)
\nonumber\\
&&
+ 19.633 g_S^4+12.362  \lambda_U g_S^2 +1.7384 \gamma g_W^2
\biggr\}, 
\la{eq:mmUnum} \\
m_H^2 &=& m_H^2(\mu)
-\left(\frac{3}{2}g_W^2+6\lambda_H+3\gamma\right)
\frac{\Sigma}{4\pi a}\nonumber\\
&&-\frac{1}{16\pi^2}\biggl\{
\left[\frac{51}{16}g_W^4+
9\lambda_H  g_W^2 
-12 \lambda_H^2+ 8 \gamma g_S^2-
3 \gamma^2\right] \biggl(\ln\frac{6}{a\mu}+\zeta\biggr) \nn
&& 
+ 4.9941 g_W^4+5.2153 \lambda_H g_W^2  + 4.6358 \gamma g_S^2
\biggr\}.
\la{eq:mmHnum}
\ea

The vacuum counterterms do not get any corrections
from $\gamma$, so that 
\ba
\langle U^\dagger U\rangle_{\rm cont} & = & 
\langle U^\dagger U\rangle_{\rm latt}-\frac{N\Sigma}{4\pi a}
-\frac{1}{(4\pi)^2}(N^2-1)g_S^2\biggl(
\ln\frac{6}{a\mu}+\zeta+\fr14 \Sigma^2-\delta
\biggr), \nn
\langle H^\dagger H\rangle_{\rm cont} & = & 
\langle H^\dagger H\rangle_{\rm latt}-\frac{\Sigma}{2\pi a}
-\frac{1}{(4\pi)^2}3g_W^2\biggl(
\ln\frac{6}{a\mu}+\zeta+\fr14 \Sigma^2-\delta
\biggr).
\la{UHpdp}
\ea
The numerical value of the constant
$\zeta+\Sigma^2/4-\delta$ is in eq.~\nr{eq:pdpnum}.

\subsection{SU(2)$\times$U(1)}
\la{sec:su2u1}

For the $g'^2$-part of standard electroweak SU(2)$\times$U(1) theory, 
only the numerical values of the lattice counterterms have been 
previously given in literature~\cite{su2u1}.
We give here the exact result in terms of the lattice 
constants $\Sigma, \zeta, \delta,\rho, \kappa_1, \kappa_4$.

The SU(2)$\times$U(1) lattice Lagrangian is 
\ba
{\mathcal L}_{\rm latt}
&=& \frac{1}{a^4 g^2}\sum_{i,j} {\rm Tr}
\Bigl[{\bf 1}-P_{ij}(x)\Bigr]+
\frac{2 \gamma^2}{a^4 g'^2} \sum_{i,j}
[1-p_{ij}^{1/\gamma}(x)] \nonumber \\
 &+& \frac{1}{a^2} \sum_{i}
\Bigl[\Tr\Phi^\dagger(x)\Phi(x)- 
\Tr\Phi^\dagger(x)U_i(x)\Phi(x+i)
e^{-i\alpha_i(x)\sigma_3} \Bigr] \nonumber \\
 &+& 
m^2\fr12\Tr\Phi^\dagger(x)\Phi(x) + \lambda
 \left[ \fr12\Tr\Phi^\dagger(x)\Phi(x)\right]^2.
\la{eq:su2u1lL}
\ea
Here $p_{ij}$ is the U(1) plaquette with the U(1) link 
$u_i(x) = \exp[{i \alpha_i(x)}]$, where 
$\alpha_i(x)=ag'B_i(x)/2$.
Any positive number $\gamma$ 
in eq.~\nr{eq:su2u1lL} gives the same naive continuum
limit. The Higgs field has been written as a matrix
$\Phi=(\tilde\phi\, \phi)$,
where $\phi$ is the standard Higgs doublet and $\tilde{\phi}
=i\sigma^2\phi^*$.
The continuum Lagrangian corresponding to eq.~\nr{eq:su2u1lL} is 
\be
L={1\over4} F_{ij}^aF_{ij}^a+{1\over4} B_{ij}B_{ij}+
(D_i\phi)^\dagger D_i\phi+m^2\phi^\dagger\phi+
\lambda(\phi^\dagger\phi)^2,
\la{eq:su2u1cL}
\ee
where $D_i=\partial_i+igA_i+ig'B_i/2$ and 
$B_{ij}=\partial_iB_j-\partial_jB_i$.

The calculation proceeds as before. The bare lattice mass
parameter is 
\ba
m^2 &=& m^2(\mu)-
\frac{\Sigma}{8\pi a}(3g^2+g'^2+12\lambda) \nn 
&-&{1\over16\pi^2}\biggl[
\biggl({51\over16}g^4-\frac{9}{8}g^2g'^2-\frac{5}{16}g'^4
+9\lambda g^2-12\lambda^2+3\lambda g'^2\biggr) 
\biggl(\ln{\frac{6}{a\mu}}+\zeta\biggr) \nn 
&+&
\frac{3}{2}g^4
\biggl(\frac{15}{32}\Sigma^2+\frac{5}{8}+\frac{\pi\Sigma}{6}
-\frac{11}{4}\delta-3\rho+2\kappa_1-\kappa_4\biggr)\nn 
&+&\frac{3}{8}g^2g'^2 \biggl(\frac{1}{4}\Sigma^2-1-2\delta\biggr)
+\frac{1}{16} g'^4
\biggl(\frac{1}{4}\Sigma^2-1-2\delta-8\rho+ 
\frac{8\pi}{3}\frac{\Sigma}{\gamma^2}
\biggr) \nn 
&+&3\lambda (3 g^2+g'^2)
\biggl( 
\frac{\Sigma^2}{4}-\delta
\biggr)\biggr]. 
\ea
The numerical value of the constant 2-loop part
on the last three  rows is
\ba
&&-{1\over16\pi^2}\biggl[
4.9941 g^4-0.88600g^2g'^2+
\Bigl(0.00932+\frac{1.66290}{\gamma^2}\Bigr)g'^4 \nn
&& \hspace*{1.5cm} +5.2153\lambda g^2+1.7384\lambda g'^2\biggr].
\la{y}
\ea
The relation of the order parameters is 
\ba
\langle\phi^\dagger\phi\rangle_{\rm cont} & = & 
\langle\phi^\dagger\phi\rangle_{\rm latt}-
\frac{\Sigma}{2\pi a} \nn 
&& -\frac{1}{16\pi^2}(3 g^2+g'^2)
\biggl(\ln{\frac{6}{a\mu}}+\zeta+\frac{1}{4}\Sigma^2-\delta\biggr).
\la{su2u1pdp}
\ea
The SU(2)+Higgs theory follows as the limit $g'\to 0$.

\subsection{U(1)}

For the U(1)+Higgs theory 
we extend  previous results~\cite{ref:hpert2} by
introducing an extra parameter $\gamma$ as 
in Ref.~\cite{su2u1} and Sec.~\ref{sec:su2u1}, 
allowing a simultaneous representation of the 
compact and non-compact lattice actions. 

The U(1)+Higgs theory  is defined by
\ba
{\mathcal L}_{\rm lat} & = & \frac{\gamma^2}{2 a^4 e^2}\sum_{i,j}
\Bigl[1-P_{ij}^{1/\gamma}(x)\Bigr] \nonumber \\
& + & \frac{2}{a^2}\sum_i\Bigl[
\phi^*(x)\phi(x)-{\rm Re}\,
\phi^*(x)U_i(x)\phi(x+i)\Bigr] \nn
& + & m^2\phi^*\phi+\lambda \Bigl(\phi^*\phi\Bigr)^2,
\ea
where $U_i(x)=\exp[iaeA_i(x)]$. 
The continuum Lagrangian is
\be
{\mathcal L}_{\rm cont}=\fr14 F_{ij}F_{ij}+
(D_i\phi)^* D_i\phi+m^2\phi^*\phi+
\lambda(\phi^*\phi)^2, \la{contlagr}
\ee
with $D_i=\partial_i+ie A_i$.
The bare mass parameter is 
\ba
m^2 & = & m^2(\mu)- 
(2e^2+4\lambda)\frac{\Sigma}{4\pi a} \nn
&& -\frac{1}{16\pi^2}\biggl[
\biggl(-4e^4+8\lambda e^2-8\lambda^2\biggr)
\biggl(\ln\frac{6}{a\mu}+\zeta\biggr) 
\nonumber \\
&& 
+e^4\biggl(\frac{1}{4}\Sigma^2
-1-2\delta-4\rho+\frac{8\pi}{3}\frac{\Sigma}{\gamma^2}\biggr)
+8\lambda e^2\biggl(\frac{1}{4}\Sigma^2-\delta\biggr)
\biggr],
\label{Ud32} 
\ea
where the numerical value of the last row is
\ba
&&
-\frac{1}{16\pi^2}\biggl[
\Bigl(-1.1068+\frac{26.6065}{\gamma^2}\Bigr)e^4+
4.6358\lambda e^2\biggr].
\ea
The relation of the order parameters is 
\be
{\langle \phi^*\phi\rangle}_{\rm cont}=
{\langle \phi^*\phi\rangle}_{\rm latt}-
\frac{\Sigma}{4\pi a}-
{e^2\over8 \pi^2}\biggl(\ln{\frac{6}{a\mu}}
+\zeta +\fr14\Sigma^2-\delta\biggr).
\la{u1pdp}
\ee
The non-compact case follows as the limit $\gamma \to\infty$.

\section{Conclusions}
\la{sec:concl}

We have applied the previously developed method of deriving
2-loop lattice-continuum relations to 3d theories relevant for
QCD and some extensions of the Standard Model at finite temperature. 
The main results are 
the expressions for the bare lattice mass parameters 
in eqs.~\nr{eq:deltam},\nr{eq:mmU},\nr{eq:mmH}, 
with numerical values as given in 
eqs.~\nr{eq:su5num},\nr{eq:mmUnum},\nr{eq:mmHnum}.
When an extrapolation to the continuum limit is made 
in the lattice simulations, these equations allow to 
extract the corresponding finite fixed value of the 
$\msbar$ mass parameter $m^2(\mu)$. Then one can compare
lattice simulations with 3d perturbation theory, 
or if $m^2(\mu)$ has been computed with methods
of dimensional reduction from 4d, one can extract the 
physical temperature to which the simulations correspond.

\section*{Acknowledgements}

We acknowledge useful discussions with K.Kajantie. M.L is grateful
to D.B\"ode\-ker for providing his computer program for SU($N$) algebra. 
M.L was partially supported by the University of Helsinki.

\appendix
\section{Feynman rules}
\label{app:verteksit}

In this appendix we give the Feynman rules for the interactions of the
Higgs field in the symmetric phase. The ones related to gauge and
ghost-gauge vertices were discussed in Sec.~\ref{sec:calc}.

In the adjoint case, the vertices can be read from the action
\begin{eqnarray}
S_\Phi&=&
\frac{1}{4}\left(\lambda_1+\frac{\lambda_2}{N}\right)\Phi^A\Phi^A\Phi^B\Phi^B
+\frac{1}{8}\lambda_2d^{ABE}d^{CDE}\Phi^A\Phi^B\Phi^C\Phi^D\nn
&&+\frac{ig}{2}\delta(p+q+r)\widetilde{(p-q)}_i
f^{ABC}\Phi^A(p)\Phi^B(q) A^C_i(r)\nn
&&+\frac{g^2}{2}\delta(p+q+r+s)(\wideundertilde{p-q})_i
f^{ACE}f^{BDE}\Phi^A(p)\Phi^B(q) A_i^C(r)A_i^D(s)\nn
&&-g^2vf^{A,D,N^2-1}f^{BCD}\overline{c}^Ac^B\Phi^C.
\la{eq:adjS}
\end{eqnarray}
In the broken phase, the shift generates two new three-point vertices
$\Phi\Phi\Phi$ and $\Phi AA$ from those in eq.~\nr{eq:adjS}, but 
there is also an additional four-gluon vertex
\begin{eqnarray}
S_A&=&-\frac{1}{24}a^2g^4v^2f^{AEH}f^{B,E,N^2-1}f^{CFH}f^{D,F,N^2-1}
A_i^AA_i^BA_i^CA_i^D.
\end{eqnarray}

In the fundamental case, correspondingly,
\ba
S_{\phi} & = & \lambda (\phi^\dagger\phi)^2 \nonumber \\
& + & g \delta(-p+q+r)\widetilde{(p+q)}_i
\phi^\dagger(p)T^A \phi(q) A^A_i(r) \nonumber \\
& + & \frac{g^2}{2}\delta(-p+q+r+s)\wideundertilde{(p+q)_i}
\phi^\dagger(p)T^{\{ A}T^{ B\}}\phi(q)
A_i^A(r)A_i^B(s)
 \nonumber \\
& + & g^2
\overline{c}^Ac^B\Bigl(
\hat{\phi}^\dagger T^A T^B \phi+\phi^\dagger T^B T^A \hat{\phi}
\Bigr). 
\ea
In the broken phase, the shift generates again 
the extra vertices $\phi \phi \phi$, 
$\phi A A$,
and there is the additional four-gluon vertex 
\be
S_A= -\frac{1}{12}a^2g^4(\hat\phi^\dagger T^A T^B T^C T^D \hat\phi)
A_i^AA_i^BA_i^CA_i^D.
\la{fundAAAA}
\ee

\section{Lattice integrals}
\label{app:ints}

In~\cite{ref:hpert1,ref:hpert2}, 
the lattice integrals were worked out for mass combinations
specific to SU(2)+fundamental Higgs. 
Let us here describe how these generalize to the present case.
Let
\ba
S(k^2,m^2) & = &  \frac{1}{\widetilde{k}^2+m^2}, \quad
V^1_{ij}(k^2,M^2) =  
\frac{\delta_{ij}}{\widetilde{k}^2+M^2}, \\
F_{ijk}(p,q,r) & = & 
\delta_{ik}\undertilde{q_k}(\widetilde{p_j-r_j})+
\delta_{kj}\undertilde{p_j}(\widetilde{r_i-q_i})+
\delta_{ji}\undertilde{r_i}(\widetilde{q_k-p_k}),
\ea
where the superscript in $V^1_{ij}$ is to remind
that we are in the Feynman gauge.
Then there are contributions from the 
following integrals~\cite{bjls}:
\ba
D^1_{\rm VVV} & = &  
\int\!dp\,dq\,
V^1_{il}(p^2,M_1^2) 
V^1_{jm}(q^2,M_2^2) 
V^1_{kn}(r^2,M_3^2) \nonumber \\
&& \hspace*{4cm} \times F_{ijk}(p,q,r)F_{lmn}(p,q,r)  \nn 
&\to &  
\frac{1}{(4\pi)^2}\sum_{i=1}^3\biggl[
-6 M_i^2\biggl(\ln\frac{6}{a\mu}+\zeta\biggr)+
\frac{1}{a}\hat{M}_i (8\pi -15\Sigma) \nn  
&+ & M_i^2 \biggl(
-\fr32+\fr54\Sigma^2-\frac{\pi}{3}\Sigma+
6\delta+6\rho-4\kappa_1+2\kappa_4
\biggr)
\biggr], \\
D^1_{\rm VVS} & = &   
\int\!dp\,dq\,
V^1_{il}(p^2,M_1^2) V^1_{jm}(q^2,M_2^2) S(r^2,m^2) 
4\delta_{ij}\undertilde{r_i}\delta_{lm}\undertilde{r_l} \nn  
&\to &  \frac{1}{(4\pi)^2}\biggl[
12 \biggl(\ln\frac{6}{a\mu}+\zeta\biggr)+2-\Sigma^2
\biggr] , \\
D^1_{\rm SSV} & = & 
\int\!dp\,dq\,
S(p^2,m_1^2) S (q^2,m_2^2) V^1_{ij}(r^2,M^2) 
(\widetilde{p-q})_i(\widetilde{p-q})_j \nn 
& \to &  
\frac{1}{(4\pi)^2}\biggl[
(M^2-2m_1^2-2m_2^2)\biggl(\ln\frac{6}{a\mu}+\zeta\biggr)-
\frac{\Sigma}{a} (\hat{m}_1+\hat{m}_2) \nn 
& + & \frac{4}{a}\hat{M} \biggl(\pi-\fr32\Sigma\biggr)+
2\delta(m_1^2+m_2^2)+4\rho M^2 \biggr], \\
D^1_{\rm GGV} & = &  
\int\!dp\,dq\,
S(p^2,m_1^2) S (q^2,m_2^2) V^1_{ij}(r^2,M^2) 
(-\widetilde{p}_i\undertilde{p_j}
\widetilde{q}_j\undertilde{q_i} ) \nn 
& \to &  \frac{1}{(4\pi )^2}\biggl[ 
\fr12 (M^2-m_1^2-m_2^2)\biggl(\ln\frac{6}{a\mu}+\zeta\biggr)+
\frac{1}{a} \hat{M} \biggl(\pi-\fr32\Sigma\biggr) \nn 
& + & \fr12 \delta (m_1^2+m_2^2)+\rho M^2\biggr],  
\la{eq:ints}
\ea
where $r=-p-q$, $\hat{m}\equiv m(1+\hat\xi a m)$,
$\hat\xi\approx 0.152859$ (however, $\hat\xi$ cancels in the final
result, see~\cite{ref:hpert1,ref:hpert2}), and
\be
\int\! dp \equiv \int_{-\pi/a}^{\pi/a}\frac{d^3p}{(2\pi )^3}.
\ee

The contributions from the figure-8 graphs follow
directly from the SU(2) case in~\cite{ref:hpert2}, 
apart from the graph (vv). Let us
note that there are two kinds of contributions
from this graph, one proportional to  
$f^{ACE}f^{BDE}
\langle A^AA^B\rangle\langle  A^CA^D\rangle$ 
(this is the isospin factor appearing in 
continuum)
and the other to the factor in eq.~\nr{eq:vvsy}.
In the former case the integral appearing is
\ba
D^{1(a)}_{\rm VV}  & = & 
\int\!dp\,dq\,
S(p^2,M_1^2) S(q^2,M_2^2) 
\biggl[12-\fr73 a^2(\widetilde{p}^2+
\widetilde{q}^2)+\frac{5}{18}a^4\widetilde{p}^2\widetilde{q}^2\biggr] \nn 
& \to & 
\frac{1}{(4\pi)^2}\sum_{i=1}^2\biggl[
\frac{1}{a} \hat{M}_i \biggl(
-12\Sigma+\frac{28}{3}\pi\biggr)+
M_i^2 \biggl(\fr73\Sigma^2-\frac{10}{9}\pi\Sigma
\biggr)
\biggr],
\ea
and in the latter case
\ba
D^{1(b)}_{\rm VV} &  = & 
\int\!dp\,dq\,
S(p^2,M_1^2) S(q^2,M_2^2) 
\biggl(-\fr19 a^4\widetilde{p}^2\widetilde{q}^2\biggr) \nn 
& \to & 
\frac{1}{(4\pi)^2} (M_1^2+M_2^2)\frac{4\pi}{9}\Sigma.
\la{eq:vvasyint}
\ea

\section{Isospin contractions}
\label{app:const}

When calculating the adjoint case mass counterterm in 
eq.~(\ref{eq:deltam})
one has to evaluate 2-loop Feynman diagrams with complicated
vertices. There are basically two ways to evaluate the isospin factors
related to the diagrams.
Either one can use the Fiertz identity
\begin{equation}
T^A_{ij}T^A_{kl}=\frac{1}{2}
\left(\delta_{il}\delta_{jk}-\frac{\delta_{ij}\delta_{kl}}{N}\right)
\end{equation}
and the projection operators in eq.~(\ref{eq:projadj}), or one can write the
fields in component form and calculate products of the structure constants
of the Lie algebra. We give here the necessary products for the latter
strategy, which usually leads to simpler algebra.

Let $T^A$ $(A\le N^2-1)$ be Hermitian traceless 
$N\times N$ matrices with
$\Tr T^AT^B=\frac{1}{2}\delta^{AB}.$ They form a representation of
the Lie algebra {\bf su}$(N)$.
Let us define $T^{N^2-1}$ according to eq.~(\ref{eq:tn2-1}).
The matrices $T^A$ can be chosen such that
the projections $P^{AB}_i$ separate the index set $I=\{1,\ldots,N^2-1\}$
to three subsets $I_1$, $I_2$ and $I_3$. The matrices corresponding to
set $I_1$ form a representation of {\bf su}$(N-1)$ and $I_3$ contains only
the index $N^2-1$. 

Since the Feynman rules for the vertices contain factors of the form
$f^{ABC}$, $f^{ABE}f^{CDE}$ and $d^{ABE}d^{CDE}$ it is obvious that
one needs to calculate sums like
\begin{eqnarray}
F_{ijk}&=&\sum_{A\in I_i}\sum_{B\in I_j}\sum_{C\in I_k}f^{ABC}f^{ABC},\\
D^{(1)}_{ijk}&=&\sum_{A\in I_i}\sum_{B\in I_j}\sum_{C\in I_k}
d^{ABC}d^{ABC},\\
D^{(2)}_{ij}&=&\sum_{A\in I_i}\sum_{B\in I_j}\sum_{C\in I}
d^{AAC}d^{BBC}.
\end{eqnarray}
Using the known results 
\begin{eqnarray}
f^{ACD}f^{BCD}&=&\delta^{AB}N,\quad
d^{ACD}d^{BCD}=\delta^{AB}\frac{N^2-4}{N},\quad
d^{AAB}=0,
\end{eqnarray}
we can deduce
the following values for $F_{ijk}$, $D^{(1)}_{ijk}$ and $D^{(2)}_{ij}$:
\begin{eqnarray}
F_{111}&=&N(N-1)(N-2),\quad
F_{122}=N(N-2),\quad
F_{223}=N,\\
D^{(1)}_{111}&=&\frac{(N+1)N(N-2)(N-3)}{N-1},\quad
D^{(1)}_{113}=2\frac{N-2}{N-1},\nn
D^{(1)}_{122}&=&N(N-2),\quad
D^{(1)}_{223}=\frac{(N-2)^2}{N},\quad
D^{(1)}_{333}=2\frac{(N-2)^2}{N(N-1)},\\
D^{(2)}_{11}&=&2N\frac{(N-2)^2}{N-1},\quad
D^{(2)}_{12}=-2(N-2)^2,\nn
D^{(2)}_{13}&=&-2\frac{(N-2)^2}{N-1},\quad
D^{(2)}_{22}=\frac{2}{N}(N-1)(N-2)^2,\nn
D^{(2)}_{23}&=&\frac{2}{N}(N-2)^2,\quad
D^{(2)}_{33}=2\frac{(N-2)^2}{N(N-1)}.
\end{eqnarray}
The results are symmetric in the permutation of the indices. All other
sums are zero.

\section{Numerical values}
\la{app:num}

Let us use the shorthand $s\equiv \sin$. Then 
the numerical constants appearing in the 2-loop 
calculation are as follows:
\ba  
\Sigma & = & \frac{1}{\pi^2}\int_{-\pi/2}^{\pi/2}\!\! d^3x
\frac{1}{\sum_i{s}^2x_i}=3.175911535625,
\la{eq:sigma} \nn
\delta & = & \frac{1}{2\pi^4}\int_{-\pi/2}^{\pi/2}d^3xd^3y
\frac{\sum_i{s}^2x_i{s}^2(x_i+y_i)}
{(\sum_i{s}^2x_i)^2\sum_i{s}^2(x_i+y_i)\sum_i{s}^2y_i}
=1.942130(1), \nn
\rho & = & \frac{1}{4\pi^4}\int_{-\pi/2}^{\pi/2}\!\! d^3xd^3y
\Biggl\{\frac{\sum_i{s}^2x_i{s}^2(x_i+y_i)}
{\sum_i{s}^2x_i\sum_i{s}^2(x_i+y_i)}-
\frac{\sum_i{s}^4x_i}
{(\sum_i{s}^2x_i)^2}\Biggr\}
\frac{1}{(\sum_i{s}^2y_i)^2} \nn
& = & -0.313964(1), \nn
\kappa_1 & = & \frac{1}{4\pi^4}\int_{-\pi/2}^{\pi/2}\!\! d^3xd^3y
\frac{\sum_i{s}^2x_i{s}^2(x_i+y_i)}
{\sum_i{s}^2x_i\sum_i{s}^2(x_i+y_i)\sum_i{s}^2y_i}
=0.958382(1), \nn
\kappa_4 & = & \frac{1}{\pi^4}\int_{-\pi/2}^{\pi/2}\!\! d^3xd^3y
\frac{\sum_i{s}^2x_i{s}^2(x_i+y_i){s}^2y_i}
{(\sum_i{s}^2x_i)^2\sum_i{s}^2(x_i+y_i)\sum_i{s}^2y_i}=
1.204295(1).
\ea
The constant $\Sigma$ can be expressed in 
terms of the complete elliptic integral of the first kind~\cite{ref:hpert1}.
The number in parentheses in the other constants 
estimates the uncertainty in the last digit. 

In \cite{ref:hpert2} two more constants $\kappa_2$ and $\kappa_3$ were
defined. However, they appear in the calculations only as a
sum $\kappa_2+\kappa_3$ and can thus be eliminated using the relation
\begin{equation}
2(\delta+\rho)=\kappa_2+\kappa_3+\kappa_4.
\end{equation}
This follows from the trigonometric identity
\begin{eqnarray}
\sin^4x+\sin^4y+\sin^4(x+y)+4\sin^2x\sin^2y\sin^2(x+y)&&\nn
-2\left(\sin^2x\sin^2y+\sin^2x\sin^2(x+y)+\sin^2y\sin^2(x+y)\right)
&=&0.
\end{eqnarray}

In addition, there is the constant $\zeta$
for which one to our knowledge 
cannot write a direct expression in momentum space:
\ba
\zeta & = & \lim_{z\to 0}\Biggl[
 \frac{1}{4\pi^4}\! \int_{-\pi/2}^{\pi/2}\!\!
\frac{d^3xd^3y}
{(\sum_i{s}^2x_i+z^2)
\sum_i{s}^2(x_i+y_i)\sum_i{s}^2y_i} -
\ln\frac{3}{z}-\fr12 \Biggr] \nn
& = & 0.08849(1).
\la{eq:zeta}
\ea
For the numerical evaluation this 
integral can be reduced to a four-dimensional one~\cite{ref:hpert1}. 
Note also that  
for the combination appearing in the relation of
lattice and continuum order parameters in 
eqs.~\nr{adjpdp},\nr{eq:pdp},\nr{UHpdp},\nr{su2u1pdp},\nr{u1pdp}, 
one gets 
\be
\zeta+\frac{1}{4}\Sigma^2-\delta = 0.66796(1).
\la{eq:pdpnum}
\ee

Finally, let us point out that 
the results above follow from a straightforward numerical integration. 
If needed, more precise values could be obtained
using the position space techniques developed in~\cite{lw}. 
We these techniques one can, for instance, write a 
closed expression for $\zeta$. Noting from
eq.~(11) in~\cite{a} that replacing the mass $am=2z$ in
eq.~\nr{eq:zeta} with an external momentum $ap$ causes 
the numerical factor $1/2$ to be replaced with $3/2$, 
one gets 
\be
\zeta = -\frac{3}{2} - \ln 6 + \sum_x 
\Bigl[(4 \pi)^2 G(x)^3 + \fr12 H(x) \Bigr]. 
\la{eq:zeta2}
\ee
Here (we have put $a=1$ as in~\cite{lw})
\be
G(x)=\int_{-\pi}^{\pi}\frac{d^3p}{(2\pi)^3}
\frac{e^{ip\cdot x}}{\widetilde{p}^2},\quad
H(x)=\int_{-\pi}^{\pi}\frac{d^3p}{(2\pi)^3}e^{ip\cdot x}
\ln{\widetilde{p}^2.}
\ee
The functions $G(x), H(x)$ 
are related by
$H(x)=2\sum_i\Bigl[
G(x)-G(x-i)
\Bigr]/({\sum_i x_i})$
for $\sum_ix_i\neq 0$, and have at $x=0$ the values
\be
G(0)=\frac{\Sigma}{4\pi}, \quad
H(0)=1.67338930297(1).
\ee
The sum in eq.~\nr{eq:zeta2}
is convergent since the 
leading asymptotic behaviours
\be
G(x)\sim \frac{1}{4\pi |x|},\quad
H(x)\sim -\frac{1}{2\pi |x|^3}
\la{eq:as}
\ee
cancel in eq.~\nr{eq:zeta2}.
To evaluate the sum one needs an efficient method of calculating
$G(x)$ very precisely~\cite{lw}.

\section{A unified representation for the gauge part of the counterterms}
\la{app:uni}

Let $F^A_{BC}=-if^{ABC}$ be the generators of the adjoint representation, 
and $G^A$ those of a general representation. We define $T$ such that
for a real (adjoint) representation, 
\be
{\rm Tr}F^AF^B=T\delta^{AB},
\ee 
and for a complex (fundamental) representation, 
\be
2{\rm Tr}T^AT^B=T\delta^{AB}.
\ee 
The quadratic Casimir $C_2$ is defined by
\be
(G^AG^A)_{ab}=C_2\delta_{ab}. 
\ee
In addition, the number of real scalar field components is denoted by $d$.
Then for the fundamental representation, 
$T=1$, $C_2=(N^2-1)/2N$ and $d=2N$, and for the adjoint representation, 
$T=C_2=N$, $d=N^2-1$. 

In the limit that the scalar self coupling constants vanish, 
one can then write the bare mass parameter in a unified way:
\begin{eqnarray}
\delta m^2&=&-2g^2 C_2\frac{\Sigma}{4\pi a} \nn
&&-\frac{g^4}{16\pi^2}C_2\Biggl[
\left(\frac{7}{2}N-3C_2-\frac{1}{2}T\right)
\left(\ln\frac{6}{a\mu}+\zeta\right)\nonumber\\
&&+\left(\frac{3}{8}N+\frac{1}{4}C_2\right)\Sigma^2
+\left(\frac{N}{2}-\frac{4}{3N}\right)\pi\Sigma
+N-C_2\nonumber\\
&&
-2\left(N+T\right)\rho
-2\left(N+C_2\right)\delta
+N\left(2\kappa_1-\kappa_4\right)\Biggr].
\end{eqnarray}
The mass-dependent part of the vacuum counterterm is 
\be
\delta V = -m^2 d \biggl[
\frac{\Sigma}{8\pi a}
+ \frac{g^2}{(4\pi)^2}C_2
\biggl(\ln\frac{6}{a\mu}+\zeta+\frac{\Sigma^2}{4}-\delta\biggr)
\biggr].
\ee

\bibliographystyle{plain}
\bibliography{contlatt}
\end{document}